\documentclass[
 reprint,
nofootinbib,
 amsmath,amssymb,
 aps,
]{revtex4-2}

\usepackage{aasmacros}
\usepackage[utf8]{inputenc}
\usepackage{graphicx}
\usepackage{tabularx}
\usepackage{dcolumn}
\usepackage{bm}
\usepackage{newtxtext,newtxmath}
\usepackage{subcaption}
\usepackage[T1]{fontenc}
\usepackage{makecell}
\usepackage{animate}
\usepackage{mathtools}
\usepackage{url}
\usepackage[hidelinks,colorlinks]{hyperref}
\usepackage{xcolor}
\hypersetup{
    colorlinks=true,
    linkcolor={blue},
    citecolor={blue},
    urlcolor={blue}
}
\begin{document}

\preprint{APS/123-QED}

\title{Calibrating the SIDM Gravothermal Catastrophe with N-body Simulations}

\author{Charlie Mace$^{1,2}$}\email{E-mail: mace.103@osu.edu}
\author{Shengqi Yang$^{3,4}$}
\author{Zhichao Carton Zeng$^{1,2,5}$}%
\author{Annika H. G. Peter$^{1,2,6,7}$}%
\author{Xiaolong Du$^{3,8}$}
\author{Andrew Benson$^{3}$}
\affiliation{$^{1}$Department of Physics, The Ohio State University, 191 W. Woodruff Ave., Columbus OH 43210, USA \\
$^{2}$Center for Cosmology and Astroparticle Physics, The Ohio State University, 191 W. Woodruff Ave., Columbus OH 43210, USA\\
$^{3}$ Carnegie Science, Observatories, 813 Santa Barbara Street,\\
$^{4}$ Los Alamos National Laboratory, Los Alamos, NM 87545, USA, \\
$^{5}$Department of Physics and Astronomy, Mitchell Institute for Fundamental Physics and
Astronomy, Texas A\&M University, College Station, TX 77843, USA \\
$^{6}$Department of Astronomy, The Ohio State University, 140 W. 18th Ave., Columbus OH 43210, USA\\
$^{7}$School of Natural Sciences, Institute for Advanced Study, 1 Einstein Drive, Princeton, NJ 08540\\
$^{8}$ Department of Physics and Astronomy, University of California, Los Angeles, CA 90095, USA\\
}

%\date{\today}

\begin{abstract}
Self-interacting dark matter (SIDM) theories predict that dark matter halos experience core-collapse in late-stage evolution, a process where the halo’s inner region rapidly increases in density and decreases in size. This process can be modeled by treating the dark matter as a gravothermal fluid, and solving the fluid equations to predict the density profile evolution. This model is incomplete without calibration to N-body simulations, through a constant factor $\beta$ included in the thermal conductivity for the long-mean-free-path limit. The value of $\beta$ employed in the gravothermal fluid formalism has varied between studies, with no clear universal value in the literature. In this work, we use the N-body code \texttt{Arepo} to conduct a series of isolated core-collapse simulations across a range of scattering cross-sections, halo concentrations, and halo masses to calibrate the heat transfer parameter $\beta$. We find that $\beta$ is independent of cross-section, halo concentration, and halo mass for velocity independent elastic scattering cross-sections.  We present a model for an effective $\beta$ as a function of a dimensionless cross-section, to describe halo evolution in the long mean free path limit, and show that it accurately captures halo evolution as long as the cross section is not too large.  This effective model facilitates comparisons between simulations and the gravothermal model, and enables fast predictions of the dark matter density profile at any given time without running N-body simulations.

\end{abstract}

\maketitle

\section{Introduction}

Self-interacting dark matter (SIDM) is a class of dark matter models that include interactions between dark matter particles \cite{spergel00,battaglieri17,tulin18,BUCKLEY18,adhikari22}. These models, which have theoretical motivation from particle physics \cite{feng2009,tulin2013a,tulin2013b,boddy2014,cline2014,FYCR2016}, may provide solutions to several unresolved contradictions between cold dark matter (CDM) models and observational data, such as the diversity problem (CDM; \cite{KuziodeNaray_2014,oman15,errani18,read19,Relatores_2019,santos20,Hayashi_2020,Li_2020}, SIDM; \cite{kaplinghat16,kamada17,sameie20,correa21,carton2022,correa22,Nadler_2023,dYang_2023, carton2023, carton2024, roberts2024}) and the cuspy halo problem (CDM; \cite{flores94,moore94,NFW97,moore99,weinberg15,10.1111/j.1365-2966.2004.07836.x}, SIDM; \cite{Kaplinghat_2020,Bhattacharyya_2022}). There are many models for these self-interactions, some with elastic scattering and some with inelastic scattering. The inelastic models produce their own distinct phenomenology \cite{chua20,oneil23,roy23}, but for this work we will focus solely on the elastic scattering case. In elastic models, the additional heat transfer channel provided by self-interactions causes the SIDM halos to undergo a time evolution not experienced by CDM halos, where an initially cuspy halo thermalizes and forms an isothermal core with a low, constant density \cite{Colín_2002}. The core formation process of an SIDM halo is followed by a slow increase in the halo's core density, a phase called core-collapse. During this phase, the core transfers energy to the outer regions of the halo, causing particles to fall into the core. This collapse, also called the gravothermal catastrophe, is a runaway process that eventually causes a rapid density increase in the center of the halo \cite{Kochanek:2000pi,balberg2002}. The phenomenon was first discovered in the study of globular clusters, where gravitational two-body scattering of stars provides the heat transfer instead of dark matter self-interaction \cite{lyndenbell1980}, but the gravothermal catastrophe model has found success in describing dark matter halos as well.

The study of SIDM core-collapse depends on reliable N-body simulations. One use of simulations is to investigate how self-interactions influence the formation of substructure  \cite{rocha2013,kahlhoefer2019,correa22,carton2022,carton2023}, but N-body simulations are also essential to the study of the simplest case of an isolated core-collapsing halo. The density profile of such a halo has been shown to follow a self-similar time evolution in the regime where the mean free path between collisions is much longer than the gravitational scale height (the long mean-free-path regime) \cite{balberg2002,koda2011,essig_SIDM,Nishikawa:2019lsc,Outmezguine:2022bhq,shengqi2023,gadnasr2023}. These density profile models, which are found from numerical solutions of gravothermal fluid equations, rely on a constant factor that must be calibrated using N-body simulations \cite{balberg2002,koda2011,essig_SIDM}. This factor is needed because the conductivity used in the long mean-free-path gravothermal fluid solution is based on the thermal conductivity of a gas, which does not include the gravitational dynamics present in a dark matter halo \cite{Outmezguine:2022bhq}. While we do not have an SIDM conductivity derived from first principles for the long mean-free-path regime, this approximation of using the thermal conductivity of a gas has been shown to be accurate up to a constant factor on the heat transfer rate \cite{balberg2002,koda2011,essig_SIDM}. This heat transfer constant, referred to as $\beta$ in this work but sometimes represented by $C$ in the literature, must be calibrated by N-body simulations.

The value of $\beta$ has been found by multiple studies, and typically is in the range $\beta\in[0.6,1.0]$ \cite{koda2011,essig_SIDM,Outmezguine:2022bhq,zhong2023}. While this may not seem like a major variation, the rapidly accelerating end of the core-collapse evolution means that the density profile of deeply collapsed structure has a strong dependence on $\beta$, which is inversely proportional to the collapse timescale in the long mean-free-path regime \cite{essig_SIDM}. The compactness of dark matter subhalos after core-collapse can have a profound effect on observable astrophysics, such as image flux ratios in substructure lensing, so a more precise understanding of $\beta$ and why its value may vary is needed.

In addition to the measured variation in $\beta$, there is also no \emph{a priori} reason to believe the value should be the same for all halos. The numerical calibration factor appears first in the treatment of globular cluster gravothermal catastrophe calculation, where the process is approximated using equations valid for heat transfer in a gas \cite{lyndenbell1980}. The calibration can be considered a ``fudge factor'' for this assumption, quantifying how close the gravothermal heat transfer rate is to the gaseous heat transfer. There is no reason to believe that this approximation is equally valid for all dark matter halos, especially as the mean free path between collisions varies with halo parameters, SIDM cross-section, and even within halos.

In this work, we propose a model for an effective calibration parameter $\beta_\mathrm{eff}$. This effective calibration includes expected heat transfer changes for halos between the short and long mean-free-path regimes, allowing our core-collapse models to apply to a broader set of scenarios. We then use N-body simulations, using the guidance of our previous work to avoid numerical errors and non-convergence \cite{mace2024convergence}, to calibrate this effective calibration across a broad range of halo and SIDM parameters.

In Section \ref{sec:methods} we introduce the framework for this study. This includes the gravothermal core-collapse model in Section \ref{sec:heatTransfer}, our $\beta_\mathrm{eff}$ model derivation in Section \ref{sec:betaModel}, N-body simulations in Section \ref{sec:sims}, and simulation analysis methods in Section \ref{sec:betaCal}. In Section \ref{sec:results} we present our simulation analysis, and we discuss implications in Section \ref{sec:conclusion}.

\section{Methods}

\label{sec:methods}
\subsection{Heat Transfer Regimes:  the Long and Short of It}
\label{sec:heatTransfer}

\begin{table*}
	\centering
	\caption{Important parameters used in this study.}
	\label{tab:params}
	\begin{tabular}{l|l}
		\hline
		Variable & Definition \\
		\hline
        \hline
        $r_s$& The scale radius for the NFW density profile, which we use as the initial condition for our simulations (Equation \ref{eqn:NFW}).\\
        \hline
        $\rho_s$& The scale density for the NFW density profile, which we use as the initial condition for our simulations (Equation \ref{eqn:NFW}).\\
        \hline
        $\sigma/m$& The SIDM scattering cross-section per unit mass, which varies between simulations.\\
        \hline
        $\hat{\sigma}$ & The dimensionless SIDM cross-section, calculated from $r_s$, $\rho_s$, and $\sigma/m$ according to Equation \ref{eqn:sighat}.\\
        \hline
        $\beta$ & The dimensionless parameter used in the lmfp heat transfer (Equation \ref{eqn:kappaLMFP}).\\
		\hline
        $\alpha$ & The dimensionless parameter connecting the lmfp and smfp heat transfer (Equation \ref{eqn:kappaTOTGEN}).\\
		\hline
        $\beta_{\mathrm{eff}}$ & The effective heat transfer parameter we define in this study, related to $\beta$ via our proposed models (Equations \ref{eqn:beta_supersimp}, \ref{eqn:beta_simp}, and \ref{eqn:betaFit}).\\
		\hline
	\end{tabular}
\end{table*}

One method of modeling SIDM core-collapse is the gravothermal fluid model, where the transport equations for a spherical, virialized,
gravothermal fluid are solved numerically, with the assumption that at each step in time the halo is in hydrostatic equilibrium. A complete description of these equations and how to solve them can be found in \cite{balberg2002}, but our challenge of core-collapse calibration can be understood through the energy flux equation:

\begin{equation}
    \frac{L(r)}{4\pi r^2}=-\kappa\frac{\partial T(r)}{\partial r},
\end{equation}
where $L(r)$ is is the luminosity through a sphere at radius $r$, and $T(r)$ is the local temperature defined by the velocity dispersion ($T(r)\propto v_{\mathrm{rms}}^2(r)$). $\kappa$ is the conductivity coefficient, which describes how effectively heat conducts through the halo at any given point in space and time. For SIDM core-collapse we assume that all of this heat conduction is from the dark matter self-interaction, but other systems (such as globular clusters) can have significant heat transfer through gravitational interactions \cite{lyndenbell1980}.

The conductivity $\kappa$ is not always calculable from first principles, but we can build a model for it by understanding limiting cases when our SIDM is more strongly or weakly interacting. First we will consider the strongly interacting regime, where the mean free path $\lambda$ of the SIDM scattering (the strength of which is determined by the cross-section per unit mass $\sigma/m$) is much smaller than the size of the system $H$ (where $H$ is approximately equal to the Jeans length $H\simeq v^2_\mathrm{rms}/G\rho$ for local mass density $\rho$ \cite{binneytremaine}). We will call this the \emph{short} mean-free-path (smfp) regime. We can define the Knudsen number $K_n\equiv\lambda/H$, and say that we are in the smfp regime when $\mathrm{K_n}\ll 1$. This means that the typical dark matter particle will scatter many times before traversing an entire orbit, and that particle's trajectory will accordingly be dominated by dark matter self-interactions rather than gravitational interactions. In the smfp regime, we can derive $\kappa_{\mathrm{smfp}}$ from first principles using kinetic theory:
\begin{equation}
    \kappa_{\mathrm{smfp}}=\frac{75\pi}{256}\frac{n\lambda^2k_B}{t_r},
    \label{eqn:kappaSMFP}
\end{equation}
with no numerical calibration required \citep{pitaevskii2012physical,essig_SIDM,shengqi2023}. In this equation $t_r$ is the local relaxation time for elastic SIDM scattering, $n$ is the local dark matter particle number density, and $k_B$ is the Boltzmann constant.

The second regime to consider is the opposite of the smfp regime, the \emph{long} mean-free-path (lmfp) regime. In the lmfp regime self-interactions are relatively weak, and $K_n\gg1$. In this case a typical dark matter particle will traverse the system many times between rare scattering events, and that individual particle's motion will be dominated by the gravitational potential rather than dark matter self-interactions. Unlike $\kappa_{\mathrm{smfp}}$, the lmfp heat conduction has not been derived from first principles. The approach in the literature (first used for globular clusters in \cite{lyndenbell1980}) has been to make an analogy with the thermal conductivity of gases, replacing the time between collisions with the relaxation time and the distance between collisions with the local Jeans length. We can write $\kappa_{\mathrm{lmfp}}$ as:
\begin{equation}
    \kappa_{\mathrm{lmfp}}=\frac{3\beta}{2}\frac{nH^2k_B}{t_r}.
    \label{eqn:kappaLMFP}
\end{equation}
This has a resemblance to Equation \ref{eqn:kappaSMFP}, but with the inclusion of the parameter $\beta$. The $\beta$ parameter is an \emph{ad-hoc} numerical parameter that describes potential deviation between the heat transfer in a gaseous system and a gravitationally bound dark matter halo.

In gravothermal collapse studies, it is common to consider either the limit $\lambda\ll H$ or $\lambda\gg H$, placing the halo firmly in the smfp or lmfp regime respectively. When a transition between the two is necessary, the total heat transfer is typically calculated as half of the harmonic mean between the two limits:

\begin{equation}
    \kappa_{\mathrm{total}}=\frac{1}{\kappa^{-1}_{\mathrm{lmfp}}+\kappa^{-1}_{\mathrm{smfp}}}.
    \label{eqn:kappaTOT}
\end{equation}
While this solution smoothly connects the two limits, there is no reason to believe the true heat transfer would  take this exact form. This transition regime prescription can be made more general by introducing a free parameter $\alpha$, and combining the conductivities as:

\begin{equation}
    \kappa_{\mathrm{total}}=(\kappa_\mathrm{lmfp}^{-\alpha}+\kappa_{\mathrm{smfp}}^{-\alpha})^{-\alpha}
    \label{eqn:kappaTOTGEN}
\end{equation}
\cite{Nishikawa:2019lsc}. We recover Equation \ref{eqn:kappaTOT} for $\alpha=1$, and as long as $\alpha>0$ this prescription satisfies the appropriate limits as $K_n$ approaches 0 or $\infty$.

Since the mean free path varies with position and time, $K_n$ cannot be defined globally for a halo. A halo may begin evolution with all particles in the lmfp regime, but as core-collapse occurs the dense central core will eventually enter the smfp regime. A common initial condition is a Navarro-Frenk-White (NFW) halo, a cuspy halo that has been shown to fit CDM N-body simulations \cite{NFW96}. The equation for an NFW density profile is:

\begin{equation}
    \rho_{\mathrm{NFW}}(r)=\frac{\rho_s}{\frac{r}{r_s}\left(1+\frac{r}{r_s}\right)^2},
    \label{eqn:NFW}
\end{equation}
where $\rho_s$ and $r_s$ are the scale density and radius parameters respectively \cite{NFW96}. While $K_n$ will vary throughout an NFW halo, we can broadly characterize the initial condition by calculating the dimensionless cross-section:
\begin{equation}
    \hat{\sigma}=(\sigma/m)\rho_s r_s
    \label{eqn:sighat}
\end{equation}
where $\sigma/m$ is the SIDM cross-section per unit mass, and $\rho_s$ and $r_s$ are the NFW scale density and length of the initial conditions \cite{essig_SIDM}. A halo with $\hat{\sigma}<1$ begins evolution with most or all of the halo in the lmfp regime, and will eventually enter the smfp regime due to core-collapse. A halo with $\hat{\sigma}>1$ begins evolution with its center in the smfp regime, and falls deeper into it due to core-collapse.

\subsection{$\beta$ Prediction}
\label{sec:betaModel}

In the lmfp regime ($\hat{\sigma}\ll1$), Equations \ref{eqn:kappaTOT} and \ref{eqn:kappaTOTGEN} both simplify to $\kappa_\mathrm{total}\simeq\kappa_\mathrm{lmfp}$. This approximation becomes less accurate for initial conditions approaching $\hat{\sigma}=1$, however, as $\kappa_\mathrm{smfp}$ has an increasing influence on $\kappa_\mathrm{total}$. If we use models calibrated to the lmfp regime (which we will in our analysis, see section \ref{sec:betaCal}), we will naturally find discrepancies in these transition regime halos, and these discrepancies could be interpreted as a variation in $\beta$. When studying how $\beta$ may vary with the SIDM cross-section and initial conditions, it is important to account for and understand this expected variation.

To quantify and investigate this effect, we define an effective parameter $\beta_{\mathrm{eff}}$, which is the value found when fitting an N-body simulation to gravothermal fluid predictions for the lmfp regime. We find $\beta_\mathrm{eff}$ for a given N-body simulation by assuming:

\begin{equation}
    \kappa_\mathrm{total}=\kappa_\mathrm{lmfp}(\beta_\mathrm{eff}).
    \label{eqn:betaeff_def}
\end{equation}
This equation is accurate in the lmfp limit for $\beta=\beta_\mathrm{eff}$ since $\kappa_\mathrm{total}\simeq\kappa_\mathrm{lmfp}$, and it may also be useful close to $\hat{\sigma}$ with an appropriately varying $\beta_\mathrm{eff}$.

To aide our analysis, we will recast our variables into dimensionless quantities. The heat conductivities (Equations \ref{eqn:kappaSMFP} and \ref{eqn:kappaLMFP}) can be written in terms of the velocity dispersion $v_{\rm rms}$ and local number density $n$ \cite{essig_SIDM}:

\begin{equation}
\begin{split}
    \kappa_\mathrm{smfp}&=\frac{2.1k_B v_\mathrm{rms}}{\sigma}\,,\\
    \kappa_\mathrm{lmfp}&=0.27\beta n v_{rms}^3 \sigma k_B / (Gm),\\
\end{split}
\end{equation}
where $G$ is Newton's gravitational constant and $k_B$ is Boltzmann's constant. For ease of calculation, we will convert the heat conduction equations to a dimensionless form $\hat{\kappa}$:

\begin{equation}
\begin{split}
    \hat{\kappa}_\mathrm{smfp}&=\frac{2.1\hat{v}_\mathrm{rms}}{\widehat{\sigma}}\,,\\
    \hat{\kappa}_\mathrm{lmfp}&=0.27\times4\pi\beta\hat{\rho}\hat{v}_\mathrm{rms}^3\hat{\sigma}\,.\\
\end{split}
\end{equation}
Here hatted variables correspond to dimensionless quantities, with $\hat{\rho}=\rho/\rho_s$ and $\hat{v}_{\mathrm{rms}}=v_\mathrm{rms}/\sqrt{4\pi G \rho_s r_s^2}$.

\subsubsection{Default heat conductivity model}
As a first approach, we will use the transition regime prescription of Equation \ref{eqn:kappaTOT} in Equation \ref{eqn:betaeff_def}, and calculate the implied $\beta_\mathrm{eff}$ as a function of $\beta$. Expanding Equation \ref{eqn:kappaTOT} gives:
\begin{equation}\label{eq:kappa_total}
        \hat{\kappa}_\mathrm{total}=\dfrac{1}{\hat{\kappa}_\mathrm{lmfp}^{-1}+\hat{\kappa}_\mathrm{smfp}^{-1}}=\dfrac{1}{\dfrac{1}{0.27\times4\pi\beta\hat{\rho}\hat{v}_\mathrm{rms}^3\hat{\sigma}}+\dfrac{\hat{\sigma}}{2.1\hat{v}_\mathrm{rms}}}.
\end{equation}
When the lmfp gravothermal solution is calibrated to an N-body simulation as described in Section \ref{sec:betaCal}, it can be thought of as approximating the ``true'' heat transfer coefficient $\beta$ with a perfectly lmfp expression. We now define the effective heat transfer as $\kappa_\mathrm{lmfp}(\beta_\mathrm{eff})= 0.27\times4\pi\beta_{\mathrm{eff}}\hat{\rho}\hat{v}_\mathrm{rms}^3\hat{\sigma}$, and relate it to the exact expression for $\kappa_\mathrm{total}$:
\begin{equation}
    \dfrac{1}{\dfrac{1}{0.27\times4\pi\beta\hat{\rho}\hat{v}_\mathrm{rms}^3\hat{\sigma}}+\dfrac{\hat{\sigma}}{2.1\hat{v}_\mathrm{rms}}}=0.27\times4\pi\beta_{\mathrm{eff}}\hat{\rho}\hat{v}_\mathrm{rms}^3\hat{\sigma}.
\end{equation}
Here again, $\beta$ is the ``true'' $\beta$ for a lmfp halo, while $\beta_{\mathrm{eff}}$ is the effective calibration value that takes into account the correction due to the lmfp to smfp transition. $\beta=\beta_{\mathrm{eff}}$ is true only at $\hat{\sigma}\rightarrow0$. We can now derive the relationship between $\beta$ and $\beta_{\mathrm{eff}}$:
\begin{equation}
\beta_{\mathrm{eff}}=\dfrac{1}{1/\beta+1.6\hat{\rho}\hat{v}_\mathrm{rms}^2\hat{\sigma}^2}\,.
\end{equation}
The halo central density and velocity dispersion values vary throughout the core-collapse process, but the halo spends most of its evolution in the cored phase. Since our quantity of interest is the time the halo takes to core-collapse, we can evaluate these parameters in the cored phase for our calculation and expect our result to give a reasonable approximation of the core-collapse time. Any inaccuracies would be in the early core formation or late collapse time periods, which are much shorter than the cored phase and will likely lead to only small errors in the total time. Even if this approximation is not accurate, if we can validate the resulting model we have a working prediction for core-collapse calibration. Using the output gravothermal solution \cite{shengqi2023}, we can calculate the halo central density and velocity dispersion at the point where the isothermal core is at its largest as $\hat{\rho}=2.2$ and $\hat{v}_\mathrm{rms}=0.30$. Using these values we have:
\begin{equation}
    \beta_{\mathrm{eff}}=\dfrac{1}{1/\beta+0.32\hat{\sigma}^2}\,.
    \label{eqn:beta_simp}
\end{equation}
Given measurements of $\beta_{\mathrm{eff}}$ from N-body simulations, we can calculate the value of $\beta$. For a simple assumption of $\beta=1$, the effective calibration is only a function of $\hat{\sigma}$:

\begin{equation}
    \beta_{\mathrm{eff}}=\dfrac{1}{1+0.32\hat{\sigma}^2}\,.
    \label{eqn:beta_supersimp}
\end{equation}
We will test this simple assumption against the case where $\beta$ is freely fit with our N-body simulations.

\subsubsection{Generalized heat conductivity model}
As mentioned previously, Equation \ref{eqn:kappaTOT} is not necessarily an accurate model for the lmfp-smfp transition regime. We will now repeat the previous calculation using the generalized transition regime prescription in Equation \ref{eqn:kappaTOTGEN}. The effective $\beta$ calibration process becomes:
\begin{equation}
    \left((3.4\beta\hat{\rho}\hat{v}_\mathrm{rms}^3\hat{\sigma})^{-\alpha}+\left(\dfrac{2.1\hat{v}_\mathrm{rms}}{\hat{\sigma}}\right)^{-\alpha}\right)^{-1/\alpha}=3.4\beta_{\mathrm{eff}}\hat{\rho}\hat{v}_\mathrm{rms}^3\hat{\sigma}\,.
\end{equation}
The relation between $\beta_{\mathrm{eff}}$ and $\beta$ becomes:
\begin{equation}
\begin{split}
    \beta_{\mathrm{eff}}&=\dfrac{1}{3.4\hat{\rho}\hat{v}_\mathrm{rms}^3\hat{\sigma}}\left((3.4\beta\hat{\rho}\hat{v}_\mathrm{rms}^3\hat{\sigma})^{-\alpha}+\left(\dfrac{2.1\hat{v}_\mathrm{rms}}{\hat{\sigma}}\right)^{-\alpha}\right)^{-1/\alpha}\\
    &=\left(\beta^{-\alpha}+\left(\dfrac{3.15}{\hat{\sigma}^2}\right)^{-\alpha}\right)^{-1/\alpha}.
\end{split}
\label{eqn:betaFit}
\end{equation}
In this study we treat $\beta$ and $\alpha$ as free parameters in our fit, and test the performance of Equation \ref{eqn:betaFit} against N-body simulations.

\subsection{Simulations}
\label{sec:sims}

We simulate isolated, dark matter only, SIDM halos using the N-body code \texttt{Arepo} \cite{arepo2010} with a previously tested SIDM module \cite{vogelsberger2012,vogelsberger13,vogelsberger14}. Our N-body simulations have three free parameters: $\hat{\sigma}$ (Equation \ref{eqn:sighat}), the initial NFW mass $M_{200m}$, and the initial NFW concentration $c_{200m}$. The subscript $200m$ here represents the mass and concentration definition used, where $M_{200m}$ is defined as the mass enclosed within the spherical region where the average density is 200 times cosmological mean mass density, and $c_{200m}$ is the concentration within that same region. $M_{200m}$ and $c_{200m}$ fully determine the halo profile just as $r_s$ and $\rho_s$ do (see Equation \ref{eqn:NFW}), and one can derive one pair of parameters from the other.

For our initial conditions, we sample 4 halo masses (ranging from $10^{7.5}\, \mathrm{M}_\odot$ to $10^{13.5}\, \mathrm{M}_\odot$) and 4 concentrations (ranging from $8$ to $26$). The masses and concentrations are chosen to cover a large range of plausible values, starting at dwarf galaxy scales and ending at scales of large galaxies comparable to the Milky Way \cite{deSalas_2019,10.1093/mnras/stw1876}. For each combination of halo mass and concentration we run isolated simulations at 6 different $\hat{\sigma}$ values (ranging from $10^{-1}$ to $10^{0.5}$) by altering the SIDM cross-section $\sigma/m$. Parameters for the full set of 96 simulations are given in Table \ref{tab:simParams}.

\begin{table*}
	\centering
	\caption{The halo parameters tested in our simulations. Each of the four subtables corresponds to one halo mass, indicated in the top left corner. Each row corresponds to one concentration, and each column to one $\hat{\sigma}$ value. The number in each cell is the SIDM cross-section $\sigma/m$ (in units of $\mathrm{cm}^2/\mathrm{g}$) required to get the desired $\hat{\sigma}$ at the indicated $M_{200m}$ and $c_{200m}$ values.}
	\label{tab:simParams}
    \begin{subtable}[t]{\textwidth}
    \centering
	\begin{tabular*}{0.7\textwidth}{@{\extracolsep{\fill}}|l||l|l|l|l|l|l|}
		\hline
		$M_{200m}=10^{7.5} \, \mathrm{M}_\odot$& $\hat{\sigma}=10^{-1}$& $\hat{\sigma}=10^{-0.75}$& $\hat{\sigma}=10^{-0.5}$& $\hat{\sigma}=10^{-0.25}$& $\hat{\sigma}=10^{0}$& $\hat{\sigma}=10^{0.5}$ \\
		\hline
        \hline
        $c_{200m}=8$ & 369.3 & 656.7 & 1,168 & 2,077 & 3,693 & 11,680 \\
		\hline
        $c_{200m}=14$ & 163.6 & 290.9 & 517.3 & 919.9 & 1,636 & 5,173\\
		\hline
        $c_{200m}=20$ & 94.49 & 168. & 298.8 & 531.4 & 944.9 & 2,988\\
		\hline
        $c_{200m}=26$ & 62.35 & 110.9 & 197.2 & 350.6 & 623.5 & 1,972\\
		\hline
	\end{tabular*}
    \end{subtable}
    \newline
    \vspace{6pt}
    \newline
    \begin{subtable}[t]{\textwidth}
    \centering
	\begin{tabular*}{0.7\textwidth}{@{\extracolsep{\fill}}|l||l|l|l|l|l|l|}
		\hline
		$M_{200m}=10^{9.5} \, \mathrm{M}_\odot$& $\hat{\sigma}=10^{-1}$& $\hat{\sigma}=10^{-0.75}$& $\hat{\sigma}=10^{-0.5}$& $\hat{\sigma}=10^{-0.25}$& $\hat{\sigma}=10^{0}$& $\hat{\sigma}=10^{0.5}$ \\
		\hline
        \hline
        $c_{200m}=8$ & 79.57 & 141.5 & 251.6 & 447.4 & 795.7 & 2,516 \\
		\hline
        $c_{200m}=14$ & 35.24 & 62.67 & 111.4 & 198.2 & 352.4 & 1,114 \\
		\hline
        $c_{200m}=20$ & 20.36 & 36.2 & 64.38 & 114.5 & 203.6 & 643.8 \\
		\hline
        $c_{200m}=26$ & 13.43 & 23.89 & 42.48 & 75.53 & 134.3 & 424.8\\
		\hline
	\end{tabular*}
    \end{subtable}
    \newline
    \vspace{6pt}
    \newline
    \begin{subtable}[t]{\textwidth}
    \centering
	\begin{tabular*}{0.7\textwidth}{@{\extracolsep{\fill}}|l||l|l|l|l|l|l|}
		\hline
		$M_{200m}=10^{11.5} \, \mathrm{M}_\odot$& $\hat{\sigma}=10^{-1}$& $\hat{\sigma}=10^{-0.75}$& $\hat{\sigma}=10^{-0.5}$& $\hat{\sigma}=10^{-0.25}$& $\hat{\sigma}=10^{0}$& $\hat{\sigma}=10^{0.5}$ \\
		\hline
        \hline
        $c_{200m}=8$ & 17.14 & 30.48 & 54.21 & 96.4 & 171.4 & 542.1 \\
		\hline
        $c_{200m}=14$ & 7.593 & 13.5 & 24.01 & 42.7 & 75.93 & 240.1 \\
        \hline
        $c_{200m}=20$ & 4.386 & 7.799 & 13.87 & 24.66 & 43.86 & 138.7\\
		\hline
        $c_{200m}=26$ & 2.894 & 5.146 & 9.151 & 16.27 & 28.94 & 91.51 \\
		\hline
	\end{tabular*}
    \end{subtable}
    \newline
    \vspace{6pt}
    \newline
    \begin{subtable}[t]{\textwidth}
    \centering
	\begin{tabular*}{0.7\textwidth}{@{\extracolsep{\fill}}|l||l|l|l|l|l|l|}
		\hline
		$M_{200m}=10^{13.5} \, \mathrm{M}_\odot$& $\hat{\sigma}=10^{-1}$& $\hat{\sigma}=10^{-0.75}$& $\hat{\sigma}=10^{-0.5}$& $\hat{\sigma}=10^{-0.25}$& $\hat{\sigma}=10^{0}$& $\hat{\sigma}=10^{0.5}$ \\
		\hline
        \hline
        $c_{200m}=8$ & 3.693 & 6.567 & 11.68 & 20.77 & 36.93 & 116.8\\
		\hline
        $c_{200m}=14$ & 1.636 & 2.909 & 5.173 & 9.199 & 16.36 & 51.73 \\
		\hline
        $c_{200m}=20$ & 0.9449 & 1.68 & 2.988 & 5.314 & 9.449 & 29.88\\
		\hline
        $c_{200m}=26$ & 0.6235 & 1.109 & 1.972 & 3.506 & 6.235 & 19.72\\
		\hline
	\end{tabular*}
    \end{subtable}
\end{table*}

Our initial halos are generated using the \texttt{SpherIC} code \cite{spherIC}. The density profile sampled by \texttt{SpherIC} is based on the NFW profile (Equation \ref{eqn:NFW}), but is suppressed at large radii to produce a finite halo mass. The sampled density profiled is equal to the NFW profile at radii less than the virial radius $R_{200}$ (defined as the radius at which the local mass density is equal to 200 times the critical density of the Universe), but for $r>R_{200}$ is truncated as:

\begin{equation}
    \rho_{\mathrm{trunc}}(r)=\rho_{\mathrm{NFW}}(R_{200})\left(\frac{r}{R_{200}}\right)^\delta e^{-\frac{10}{7}\left(\frac{r}{R_{200}}-1\right)}
\end{equation}
with the cutoff slope:
\begin{equation}
    \delta = {\frac{10}{3}-\frac{1+3c_{200m}}{1+c_{200m}}}.
\end{equation}

Based on our findings in \cite{mace2024convergence}, we run all simulations with $N=10^6$ particles, $\eta=0.001$ (where $\eta$ is the dimensionless parameter dictating the timestep size), and softening length $\epsilon$ given by

\begin{equation}
    \epsilon(N,c) = r_s\left[\ln{(1+c) - \frac{c}{1+c}}\right]\sqrt{\frac{0.32(N/1000)^{-0.8}}{1.12c^{1.26}}}
    \label{eqn:soft}
\end{equation}
\cite{vdB2018,carton2022}, as these choices resulted in well converged results. Each simulation is run into the core-collapse regime, where the central density is at least $100\rho_s$.

\begin{figure}[h]
\begin{subfigure}[t]{\columnwidth}
    \centering
	\includegraphics[width=\textwidth]{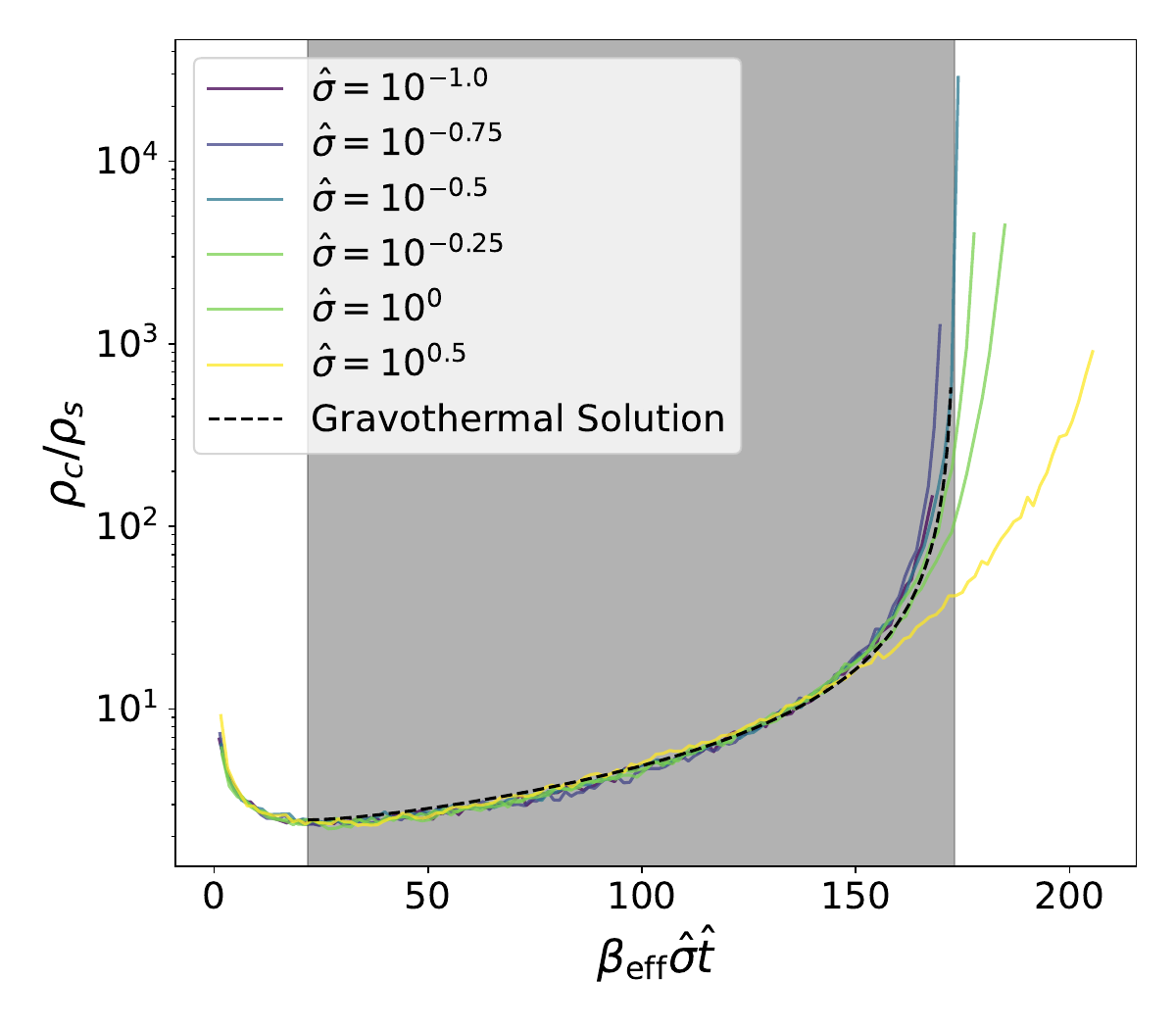}
\end{subfigure}
    \caption{Halo central density evolution for the simulation batch with $M_{200m}$. The horizontal axis is a dimensionless time unit scaled by the best fit $\beta_\mathrm{eff}$ found for each individual simulation. Each line color corresponds to one $\hat{\sigma}$ value as noted in the legend, and the black dashed line is the gravothermal solution (Equation \ref{eqn:shengqi_rhoc}).}   \label{fig:betaFits}
\end{figure}

\begin{figure*}
	\includegraphics[width=2\columnwidth]{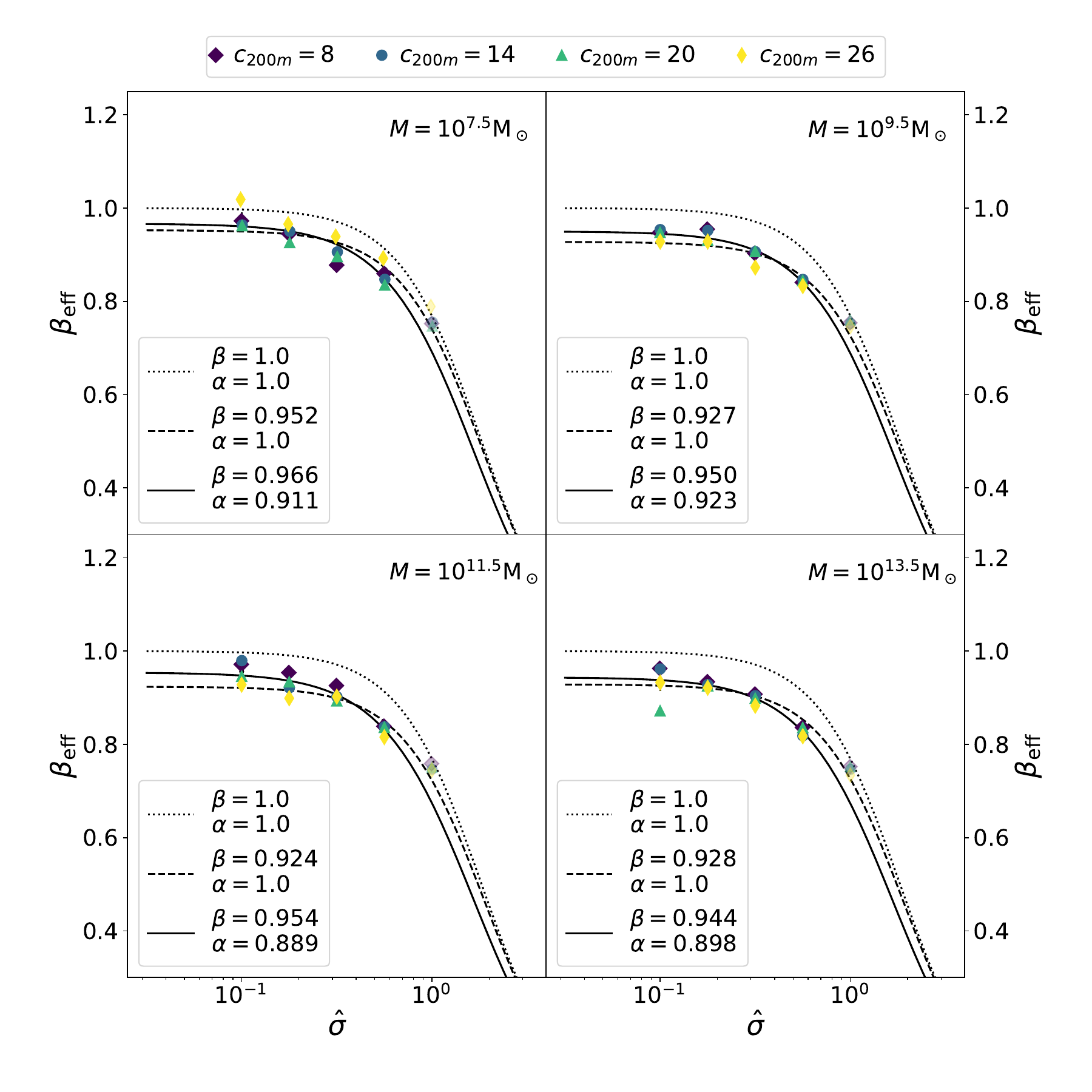}
    \caption{Calibration parameter $\beta_{\mathrm{eff}}$ as a function of $\hat{\sigma}$ for four different initial halo masses. For each mass, the halo concentration is displayed by the marker color and symbol. The $\hat{\sigma}=1$ data, which was not included in the model fitting, is displayed in lower opacity. The lines represent our three models. The dotted black line is the simplest case of Equation \ref{eqn:beta_supersimp}, with $\beta=\alpha=1$ and no free parameters. The dashed black line is the fit to Equation \ref{eqn:beta_simp}, where $\beta$ can vary and $\alpha=1$. The solid black line is the fit to Equation \ref{eqn:betaFit}, Where both $\beta$ and $\alpha$ are permitted to vary. Best fit values for our models are given in the panel legends.} \label{fig:beta_4panel}
\end{figure*}

\subsection{Measuring $\beta$}
\label{sec:betaCal}

To measure $\beta$ from an N-body simulation, we first compute the halo central density as a function of time. To calculate the central mass density for a given snapshot, we use the same method as \cite{mace2024convergence}, which is based on \cite{koda2011}. This calculation assumes the profile within the core radius $r_c$ is an isothermal sphere, and finds $r_c$ iteratively using the constraint:

\begin{equation}
    r_c=\sqrt{v_c^2/4\pi G\rho_c}
    \label{eqn:CD_cond}
\end{equation}
with $v_c$ as the velocity dispersion inside $r_c$, and the density $\rho_c$ defined as a function of the mass $M(r_c)$ enclosed within $r_c$:
\begin{equation}
    \rho_c\equiv 1.10\times M(r_c)\left/\frac{4}{3}\pi r_c^3\right. .
    \label{eqn:centralDensity}
\end{equation}
The factor of 1.10 is the ratio between the central density and the average density within $r_c$, and can be found in \cite{koda2011}. To calculate $r_c$ we begin with an initial guess and use that value to compute $v_c$ and $\rho_c$. We then use those values in Equation \ref{eqn:CD_cond} to find a new $r_c$, and iterate until the radius converges.

Once the central density evolution is computed, we measure $\beta_\mathrm{eff}$ by fitting our simulation data to the predicted evolution found in \cite{shengqi2023}:
\begin{equation}
\begin{split}
    \log{\rho_{\mathrm{core}}}=&A[\log{\beta_{\mathrm{eff}}\hat{\sigma}\hat{t}-(E+3)}]^2 + C \\
    &+ \frac{D}{E+0.0001-\log{\beta_{\mathrm{eff}}\hat{\sigma}\hat{t}}}
    \label{eqn:shengqi_rhoc}
\end{split}
\end{equation}
where A, C, D, and E are constants calibrated to a gravothermal fluid solution in the lmfp regime, and $\hat{t}=t\sqrt{4\pi G\rho_s}$. We use $A=-0.1078$, $C=-21.64$, $D=21.11$, and $E=2.238$ \cite{shengqi2023}. This model is valid from core formation until collapse, when $1.341<\log_{10}(\beta_{\mathrm{eff}}\hat{\sigma}\hat{t})<2.238$, and we consider only this region of data in our fitting.

The gravothermal fluid solution the above model is calibrated to was run with 150 radial bins. Preliminary convergence tests with up to 300 bins showed small deviations in the core-collapse evolution, suggesting there may be numerical errors in the solution due to non-convergence. Since the difference was small and increasing the bin count was numerically costly, we are proceeding in our analysis using the 150 bin count and investigating the potential numerical effects further for use in future work.

The model above is calibrated to the lmfp regime, but in our study we simulate halos up to $\hat{\sigma}=10^{0.5}$. This allows us to test the transition region between long and short mean-free-paths, and determine the range of validity for the lmfp solution. 

\section{Results}

\label{sec:results}

\begin{figure}
\begin{subfigure}[t]{\columnwidth}
    \centering
	\includegraphics[width=\textwidth]{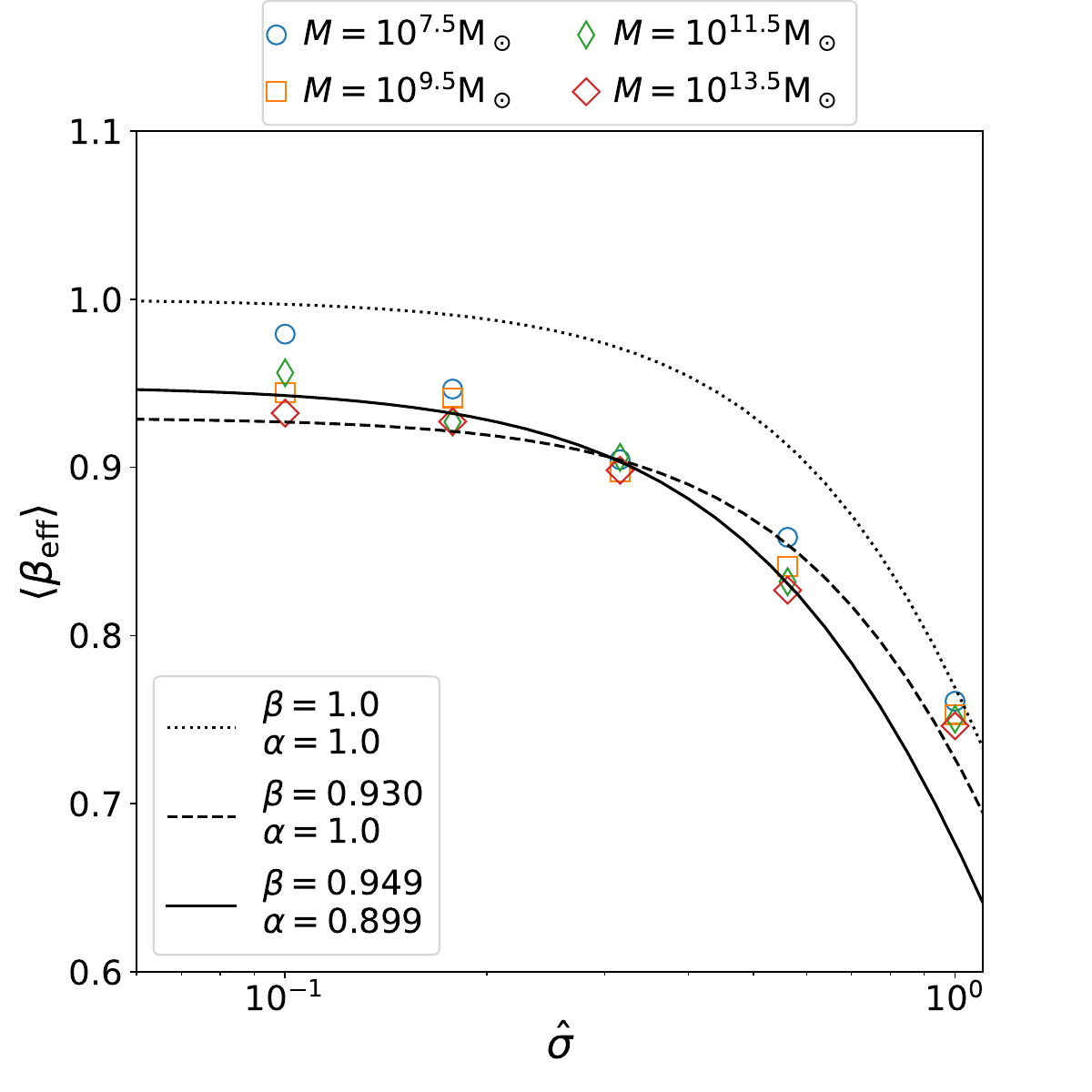}
\end{subfigure}
    \caption{Calibration parameter $\beta_{\mathrm{eff}}$ averaged over initial halo concentration. The solid line is the best fit Equation \ref{eqn:betaFit}, with fit values $\beta$ and $\alpha$ displayed in the top right corner. The dashed black line is the fit to Equation \ref{eqn:beta_simp}, which is the $\beta_\mathrm{eff}$ prediction with $\alpha=1$. The dotted black line is the simplest case, with $\beta=\alpha=1$ and no free parameters.}   \label{fig:betaFull}
\end{figure}
Figure \ref{fig:betaFits} shows the $\beta_{\mathrm{eff}}$ fitting results for the simulation set with $M_{200m}=10^{10.5}\mathrm{M}_\odot$ and $c_{200}=26$. We choose this case as an example, of both the core-collapse process and our $\beta_\mathrm{eff}$ fitting process. In this figure we plot the dimensionless central density (calculated with Equation \ref{eqn:centralDensity}) against a dimensionless time parameter scaled by $\beta_\mathrm{eff}$. In this figure we can see the core-collapse process: the density first falls to a minimum in the cored phase, and then slowly increases until suddenly and sharply increasing as the core collapses.

Following Section \ref{sec:betaCal}, the best fit $\beta_{\mathrm{eff}}$ for each individual simulation is found by fitting the central density (solid lines, $\hat{\sigma}$ designated by color) to Equation \ref{eqn:shengqi_rhoc} (dashed black line). The only effect of changing $\beta_\mathrm{eff}$ in this figure is scaling the horizontal axis for an individual simulation, which will align that simulation to the gravothermal solution at the correct $\beta_\mathrm{eff}$. For this fit we only consider data after core formation and before critical core collapse, by the criterion $1.341<\log_{10}(\beta_{\mathrm{eff}}\hat{\sigma}\hat{t})<2.238$. This is the effective range of Equation \ref{eqn:shengqi_rhoc}, and can be seen as the shaded region in Figure \ref{fig:betaFits}.

We can see that for this sample of simulations, the fit is reasonably accurate (predicting evolution close to the model up until the final rapid stage of core-collapse) for all $\hat{\sigma}$ values except for the largest, $\hat{\sigma}=10^{0.5}$, which visibly deviates from the gravothermal model earlier and more significantly than the other simulations. The deviation of the largest $\hat{\sigma}$ is not surprising, since it is the simulation which is most strongly in the smfp regime, while Equation \ref{eqn:shengqi_rhoc} is known only to be valid in the lmfp regime. 

Since $\hat{\sigma}=10^{0.5}$ is not well fit by the gravothermal model, the $\beta_{\mathrm{eff}}$ measured from these simulations will not be robust, and different values could be found by altering the way in which the model is fit to the data. For this reason we will not include the $\beta_{\mathrm{eff}}$ values measured from the smfp simulations in the rest of our results, and instead focus on $\hat{\sigma}\leq1.0$ where the fit is accurate. Figure \ref{fig:betaFits} is an example of only one mass and concentration pair, but all other combinations have a similar fit accuracy.

Once $\beta_\mathrm{eff}$ values are measured for all simulations, we use the \texttt{optimize.curve\_fit} method \cite{curvefit} from the \texttt{scipy} Python package \cite{scipy} to fit Equations \ref{eqn:beta_simp} (where $\alpha=1$ and $\beta$ varies) and \ref{eqn:betaFit} (where both $\beta$ and $\alpha$ can vary) to the data. For this fitting, we omit the data at $\hat{\sigma}=1$, because neither of our models were capable of accurately representing it simultaneously with the lowest $\hat{\sigma}$ data. Since we are using the lmfp heat transfer model to construct these fits, it is unsurprising that reaching $\hat{\sigma}$ may produce behavior we cannot accurately represent.

Figure \ref{fig:beta_4panel} shows $\beta_{\mathrm{eff}}$ measurements as a function of $\hat{\sigma}$ for all masses and concentrations.  For each mass we display the Equation \ref{eqn:betaFit} best fit as a solid black line, and include the fit values panel legends. We also display the fit to Equation \ref{eqn:beta_simp} (which fixes $\alpha=1$ but allows $\beta$ to vary, as is often done in the literature \cite{balberg2002,essig_SIDM}) for each panel as the dashed black line, and display fit values in the legends. The dotted black line on each panel is Equation \ref{eqn:beta_supersimp} where $\beta = \alpha = 1$, our simplest case using the order of magnitude estimate for $\beta$. All fits assume $\beta$ and $\alpha$ are constant for each halo mass, and do not vary with concentration or $\hat{\sigma}$.

To evaluate the accuracy of our models, we compute the reduced chi-squared test statistic $\chi^2_\nu$ for each combination of model (Equation \ref{eqn:beta_supersimp}, \ref{eqn:beta_simp}, or $\ref{eqn:betaFit}$) and halo mass. For our analysis we assume the dominant error is the scatter in $\beta_{\mathrm{eff}}$ between different $c_{200m}$ values, quantified as the standard error of the mean. We report $\chi^2_\nu$ values for the Figure \ref{fig:beta_4panel} fits in the first four rows of Table \ref{tab:stats}. These statistics show that Equation \ref{eqn:betaFit} (where we allow both $\beta$ and $\alpha$ to vary as fit parameters) is the preferred model in each case (as this model has $\chi^2_\nu$ closest to 1), although the highest mass may be overfitting.

Qualitatively, Equations $\ref{eqn:beta_simp}$ and $\ref{eqn:betaFit}$ appear to behave similarly with some small noticeable differences. Equation \ref{eqn:beta_simp}, where $\alpha$ is fixed at 1, matches the $\hat{\sigma}=1$ data well even though these points were not included in the fit. This fit does not predict lower $\hat{\sigma}$ data well however, leveling out at a $\beta_\mathrm{eff}$ value slightly lower than the data. Equation \ref{eqn:beta_simp}, where $\alpha$ is allowed to vary, shifts to match the low $\hat{\sigma}$ data at the expense of $\hat{\sigma}=1$.

\begin{table}
	\centering
	\caption{$\chi^2_\nu$ values for the fits performed in Figure \ref{fig:beta_4panel} (top four rows) and Figure \ref{fig:betaFull} (bottom row).}
	\label{tab:stats}
	\begin{tabular}{|l||l|l|l|}
		\hline
        & Eqn. \ref{eqn:beta_supersimp} & Eqn. \ref{eqn:beta_simp} & Eqn. \ref{eqn:betaFit} \\
        \hline
        \hline
        $M_{200m}=10^{7.5} \mathrm{M}_\odot$ & $20.6$ & $3.11$ & $2.03$ \\
        \hline
        $M_{200m}=10^{9.5} \mathrm{M}_\odot$ & $204$ & $12.0$ & $1.28$ \\
        \hline
        $M_{200m}=10^{11.5} \mathrm{M}_\odot$ & $92.0$ & $6.85$ & $0.535$ \\
        \hline
        $M_{200m}=10^{13.5} \mathrm{M}_\odot$ & $271$ & $10.9$ & $0.0321$ \\
        \hline
        \hline
        All Masses & $335$ & $6.97$ & $1.71$ \\
        \hline
	\end{tabular}
\end{table}

There is no clear correlation between $\beta_\mathrm{eff}$ and concentration, with the exception of the $c=26$ case appearing on the extreme end of the range for a given $\hat{\sigma}$. This suggests that $\beta_{\mathrm{eff}}$ does not significantly depend on concentration, and that the observed variation may be due to realization noise in the initial conditions or scattering calculations.

Considering the possibility that the $\beta_{\mathrm{eff}}$ variation across concentration values may not be physically meaningful, we can calculate the average calibration factor $\langle\beta_{\mathrm{eff}}\rangle$ across concentration at each $M_{200m}$ and $\hat{\sigma}$ combination. Figure \ref{fig:betaFull} shows the variation of this concentration-averaged $\beta$ as a function of $\hat{\sigma}$, with different halo masses indicated with marker style. We also fit both Equations \ref{eqn:beta_simp}, \ref{eqn:betaFit} with the same method as in Figure \ref{fig:beta_4panel}, and display the simple $\beta=\alpha=1$ model of Equation \ref{eqn:beta_supersimp}.

The fits in Figure \ref{fig:betaFull} perform similarly to the individual mass fits in Figure \ref{fig:beta_4panel}, and $\chi^2_\nu$ values are given in the final row of Table \ref{tab:stats}. We see that Equation \ref{eqn:beta_supersimp} (dotted black line, $\beta=\alpha=1$) fails to fit the data, but that the generalized heat conductivity (solid black line, Equation \ref{eqn:betaFit}) works well for $\hat{\sigma}<1$. The intermediate model with $\alpha=1$ but $\beta$ allowed to vary finds has similar drawbacks as in Figure \ref{fig:beta_4panel}, with difficulty fitting both the asymptotic $\hat{\sigma}\ll1$ and the bend in measured $\langle\beta_\mathrm{eff}\rangle$ simultaneously.

\section{Summary and Discussion}

\label{sec:conclusion}

In this study we investigate the SIDM core-collapse calibration factor $\beta$, and its potential variation between different dark matter halos and SIDM cross-sections. We find that the gravothermal model matches N-body simulation results if we treat $\beta$ as a constant value, but define an effective calibration $\beta_\mathrm{eff}$ that accounts for expected variation in the heat transfer as a function of halo parameter $\hat{\sigma}$. We propose two methods for modeling $\beta_\mathrm{eff}$: Equation \ref{eqn:beta_simp} which includes $\beta$ as the only fit parameter, and Equation $\ref{eqn:betaFit}$ which also allows the transition regime slope $\alpha$ to vary. We find that:

\begin{itemize}
    \item The core-collapse evolution in our N-body simulations is accurately modeled by the lmfp gravothermal model (Equation \ref{eqn:shengqi_rhoc}) when we treat $\beta$ and $\alpha$ as halo independent parameters, and compute the $\beta_\mathrm{eff}$ parameter as a function of only $\hat{\sigma}$. This method is effective for $\hat{\sigma}<1$ (see Figure \ref{fig:betaFull}).
    \item Our two models for $\beta_\mathrm{eff}$ (Equations \ref{eqn:beta_simp} and $\ref{eqn:betaFit}$) perform substantially better than the order of magnitude approximation (Equation \ref{eqn:beta_supersimp}), and allowing $\alpha$ to vary increases performance at low $\hat{\sigma}$ compared to fixing $\alpha=1$.
\end{itemize}

The success of our model for halo independent $\beta$ and $\alpha$ values suggests that some of the variation of $\beta_{\mathrm{eff}}$ values used in the literature is due to inaccurate heat transfer in the lmfp to smfp transition regime, rather than a variation in the original $\beta$ parameter derived for the lmfp regime in \cite{lyndenbell1980}. If a $\beta$ fitting method assumes the lmfp heat transfer, the value being found is what we define as $\beta_\mathrm{eff}$. Regardless of the reason for the variation we find, calibrating a single $\beta_{\mathrm{eff}}$ value and attempting to apply it to different halos will lead to inaccurate results, especially in measurements that are sensitive to the core-collapse time, like substructure lensing.

While the calibration variation quantified in this study can account for some of the different $\beta$ values in the literature, it cannot account for all of it. For example, the N-body simulation of an isolated, core-collapsing, NFW halo presented in \cite{koda2011} and further analyzed in \cite{essig_SIDM} reports $\beta=0.75$. This halo has parameters $M_{200m}=10^{10.56}$, $c_{200m}=9.20$, and $\hat{\sigma}=0.088$, placing it well within the range of parameters we tested. Using our best fit parameters in Equation \ref{eqn:betaFit}, we would predict $\beta_\mathrm{eff}=0.949$. When we apply our $\beta_\mathrm{eff}$ fitting method to the N-body data used in \cite{koda2011}, we find $\beta_\mathrm{eff}\simeq0.7$, which clearly contradicts our predicted value\footnote{N-body data acquired in private communication with the authors of \cite{koda2011}.}. Our simulations were conducted with the convergence guidelines developed in our previous study \cite{mace2024convergence}, suggesting there is a significant different between the simulation in \cite{koda2011} and the simulations conducted in this study. This difference could be initial conditions, boundary conditions, or any number of simulation parameters and methods that differ between \texttt{Arepo} and the version of \texttt{GADGET} adapted for use in \cite{koda2011}.

In addition to the potential systematic effects from halo initial conditions, boundary conditions, or simulation parameters, there is also further work to be done of the accuracy of calibrating $\beta_\mathrm{eff}$ with our method. We assume the scatter in $\beta_\mathrm{eff}$ values between our simulations at different concentration values to be the dominant source of uncertainty when computing the statistics in Table \ref{tab:stats}, but we do not account for initial condition realization noise or uncertainty in measuring halo central density. We run simulations with $10^6$ to try and avoid significant realization noise as indicated in our convergence testing study \cite{mace2024convergence}, but the exact level of variation from this effect is unclear without sampling more halos from the specific set of initial condition parameters used in this paper. Finally, we assume that the scatter in $\beta_\mathrm{eff}$ values between different halo concentrations and masses is attributable to random variation, but it is possible there is a real systematic dependence on those parameters rather than just $\hat{\sigma}$. While that variation appears to be small enough to ignore in this study, future more precise observational measurements may demand equally precise analytical methods, and our method may require adaptation.

Our results support a fixed value $\beta$ and $\alpha$ for all halos with $\hat{\sigma}<1$, regardless of mass, concentration, or SIDM cross-section. The $\beta_\mathrm{eff}$ model of Equation \ref{eqn:betaFit} with $\beta=0.949$ and $\alpha=0.899$ matches the N-body data simulated across a wide range of halo concentrations and masses, suggesting our findings can be applied to a diverse set of dark matter halos. This model for $\beta_\mathrm{eff}$ can be used in the empirical density profile discussed in \cite{shengqi2023} to predict the state of a core-collapsing halo as a function of time, without the need to calibrate to costly N-body simulations.

\section*{Acknowledgements}

We thank Frank van den Bosch, Stacy Kim, Daneng Yang, and Manoj Kaplinghat for useful discussions.

This work was supported in part by a grant from the OSU College of Arts and Sciences Division of Natural and Mathematical Sciences. This work was also supported by grant NSF PHY-2309135 to the Kavli Institute for Theoretical Physics (KITP). S. Yang acknowledges support from the Director’s postdoctoral fellowship funded by the Laboratory Directed Research and Development (LDRD) program of Los Alamos National Laboratory (LANL) under project No. 20240863PRD2. Z.C. Zeng is partially supported by the Presidential Fellowship of the Ohio State University Graduate School.

\bibliography{apssamp}

\end{document}